\documentclass[twocolumn,showpacs,preprintnumbers,aps,amsmath,amssymb,
floatfix,superscriptaddress]{revtex4}
\usepackage[T1]{fontenc} 
\usepackage{graphicx,epsfig,longtable}
\usepackage{epstopdf} 
\usepackage{amssymb}
\usepackage{amsmath}
\usepackage[english]{babel}
\usepackage{color}  
\usepackage[pdftex,colorlinks,linkcolor = blue,urlcolor  = blue,citecolor = blue]{hyperref}

\usepackage{dcolumn,booktabs}
\newcolumntype{d}[1]{D{.}{.}{#1}}

\usepackage{lipsum}

\definecolor{mbscolor}{rgb}{0.60, 0.0, 0.65}

\usepackage{dsfont}

\def\ga{\,\,\raise0.14em\hbox{$>$}\kern-0.76em\lower0.28em\hbox
{$\sim$}\,\,}
\def\la{\,\,\raise0.14em\hbox{$<$}\kern-0.76em\lower0.28em\hbox
{$\sim$}\,\,}

\begin{document}


\title{Comprehensive test of nuclear level density models}

\author{Stephane Goriely} 
\email{stephane.goriely@ulb.be}
\affiliation{Institut d'Astronomie et d'Astrophysique, Universit\'e Libre de Bruxelles, Campus de la Plaine CP 226, 1050 Brussels, Belgium}
\author{Ann-Cecilie Larsen}
\email{a.c.larsen@fys.uio.no}
\affiliation{Department of Physics, University of Oslo, N-0316 Oslo, Norway}
\author{Dennis~M\"{u}cher}
\email{dmuecher@uoguelph.ca}
\affiliation{College of Physics \& Engineering Science, University of Guelph, 50 Stone Road East Guelph, Ontario N1G 2W1, Canada}
\affiliation{TRIUMF, 4004 Wesbrook Mall, Vancouver, British Columbia, V6T 2A3, Canada}

\date{\today}

\begin{abstract} 
For the last two decades, experimental information on nuclear level densities for about 60 different nuclei has been obtained on the basis of the Oslo method.
While each of these measurements has been typically compared to one or a few level density models, a global study including all the measurements has been missing. 
The present study provides a systematic comparison between Oslo data and six global level density models for 42 nuclei for which $s$-wave resonance spacings are also available. 
We apply a coherent normalization procedure to the Oslo data for each of the six different models, all being treated on the same footing. 

 Our  quantitative analysis shows that the constant-temperature model presents the best global description of the Oslo data, closely followed by the mean-field plus combinatorial model and Hartree-Fock plus statistical model. Their accuracies are quite similar, so that it remains difficult to clearly favour one of these models. When considering energies above the threshold where the experimental level scheme is complete, all the six models are shown to lead to rather similar accuracies with respect to Oslo data.

The recently proposed shape method can, in principle, improve the situation since it provides an absolute estimate of the excitation-energy dependence of the measured level densities. 
We show for the specific case of $^{112}$Cd that the shape method could exclude the Hartree-Fock plus statistical model. 
Such an analysis remains to be performed for the bulk of data for which the shape method can be applied to the Oslo measurements before drawing conclusions on the general quality of a given nuclear level density model. 

\end{abstract}


\maketitle

\section{Introduction}
\label{sec:intro}

Nuclear level densities (NLDs) play a key role in basic nuclear-physics research as well as in many applications. 
The study of NLDs has been an active field of research from Bethe's pioneering work in 1936 \cite{Bethe36}. 
Based on Bethe's Fermi gas model, a large number of analytical formulas have been proposed to describe not only the exponential increase of levels with excitation energy, but also the impact of shell, pairing and collective effects (see, \textit{e.g.}, Ref.~\cite{Capote09} and references therein).

Level densities are required for modelling nuclear reactions when the number of available quantum levels is too large for a level-by-level description to be meaningful. 
With the development of next-generation experimental facilities for radioactive ion beams, as well as for astrophysical purposes, nuclear data far away from the valley of stability are required. 
This poses a huge challenge for NLD models. 
Indeed, cross section predictions have mostly relied on more or less phenomenological approaches, depending on parameters adjusted to scarce experimental data for nuclei close to the valley of $\beta$-stability, or deduced from systematics. 
Such an approach is expected to be reliable for nuclei not too far from experimentally accessible regions, but are questionable when dealing with exotic nuclei. 
To face such difficulties, it would be preferable to rely on methods that are as fundamental (microscopic) as possible and based on physically sound models, and additionally that can be applied systematically to the large number of nuclei of interest in nuclear applications.

Microscopic models of NLDs have been developed for the last decades (see e.g. \cite{Goriely08b,Hilaire12,Alhassid15,Dossing19,Fanto21} and references  therein), but they are seldom used for practical applications. 
This is often due to their lack of accuracy in reproducing experimental data (especially when considered globally on data for many nuclei), or their determination of only a limited number of nuclei, or because they do not offer the same flexibility (parameter adjustment) as analytical expressions with tuneable parameters. 
The combinatorial approach proposed in Refs. \cite{Goriely08b,Hilaire12} demonstrated that such models can compete with the phenomenological ones in the global reproduction of experimental data and that local adjustment of the tabulated NLDs can be obtained with simple analytical corrections. 
This approach provides energy-, spin-, and parity-dependent NLDs that, at low energies, describes the non-statistical limit. 
This limit cannot, by definition, be described by the traditional, statistical formulae. 
Such a non-statistical behaviour can have a significant impact on cross section predictions, particularly when calculating cross sections sensitive to spin and/or parity distributions, such as isomeric production cross sections or capture cross sections \cite{Goko06}. 
However, the combinatorial method also needs improvement because of the phenomenological aspects of some of its ingredients, which hampers its microscopic nature, and consequently its predictive power. 

When considering global and publicly accessible NLD models providing predictions for a large number of nuclei, only a limited number of methods are available. 
These include variants of the Fermi-Gas model \cite{Capote09,Koning08}, the statistical model based on a mean-field single-particle level scheme and pairing properties \cite{Demetriou01}, and the combinatorial approach of Refs. \cite{Goriely08b,Hilaire12}. 
In particular, a collection of NLD tables is publicly available in the TALYS reaction code \cite{Koning12}. 
Each of these models has its weaknesses and strengths.  
It remains difficult to favour one specific approach, since they all more or less reproduce equally well the overall set of NLD experimental data that essentially consist of the low-lying levels and the $s$-wave resonance spacings at the neutron separation energy.
While the standard Fermi gas model predicts the NLD to be proportional to the exponential of the entropy, hence of the square root of the excitation energy, the inclusion of a low-energy constant-temperature behaviour, or different variants of the energy-dependent shell, pairing or collective effects in each of these models give rise to different dependences of the NLD with the excitation energy  \cite{Capote09}.
Since early 2000, new experimental data based on the Oslo method has been made available \cite{Guttormsen87,Guttormsen96,Schiller00,Oslo}, and consists today in a relatively large set of data including some 60 different nuclei. 
While each of these measurements have been typically compared to one or a few NLD models, a global study including all the measurements has not been performed. 
Consequently, it remains  to be seen if such a comprehensive set of measurements can provide some insight on the validity or performance of the NLD models for a large range of nuclei at excitation energies below the neutron
 separation energy. 
This is the objective of the present study.

Recently, Wiedeking et al.~\cite{Wiedeking21} proposed an upgrade of the Oslo method, the so-called  ``shape method'' to estimate the slope of the $\gamma$-ray strength function (GSF) extracted from the Oslo measurements. 
Within the Oslo method, the slope of the GSF is directly connected to the slope (and thus the energy dependence) of the experimentally extracted NLD. Recently, M\"ucher et al.~\cite{Mucher22} demonstrated that, by combining the Oslo method and the shape method, the partial absolute NLD can be extracted, experimentally. NLD models were tested against data for two even-even nuclei. In this case the NLD only needs to be normalized to the low-lying discrete levels. This essentially model-free approach consequently presents an important advantage with respect to the Oslo method, as it does not need to normalize the NLD on neutron resonance data at the neutron 
 separation energy. 

In Sec.~\ref{sec:meth}, the methodology is presented, with a specific description of the six global NLD models considered and the proposed renormalization procedure. 
In Sec.~\ref{sec:test}, the renormalization procedure is illustrated on four nuclei to emphasize the differences obtained when adopting different NLD models. 
In Sec.~\ref{sec:sys}, the same procedure is applied to the 42 nuclei for which Oslo measurements have been performed and for which experimental $s$-wave resonance spacings are available. 
The accuracy of the excitation-energy dependence predicted by each of the six NLD formulae is quantitatively deduced by this analysis. 
In Sec.~\ref{sec:shape}, the new shape method is considered and the model-free excitation-energy dependence of the NLD is extracted for one nucleus, $^{112}$Cd to test the model predictions. 
Finally, conclusions are drawn in Sec.~\ref{sec:con}.

\section{Methodology}
\label{sec:meth}

\subsection{NLD models}
\label{sec:mod}

Six global and publicly available NLD models, as included in the TALYS reaction code \cite{Koning12}, are considered in the present study. These include
\begin{itemize}
\item the Constant-temperature model~\cite{Ericson59,Gilbert65} combined with the Fermi gas model~\cite{Bethe36}, hereafter referred to as Cst-T \cite{Koning08}, 
\item the Back-shifted Fermi gas model (BSFG) \cite{Koning08}, 
\item the Generalized superfluid model (GSM) \cite{Ignatyuk79,Capote09}, 
\item the Skyrme-Hartree-Fock plus statistical (HF+stat) model \cite{Demetriou01} 
\item the Skyrme-Hartree-Fock-Bogoliubov plus combinatorial method (HFB+comb) \cite{Goriely08b}
\item the  temperature-dependent Gogny-Hartree-Fock-Bogoliubov plus combinatorial method (THFB+comb) \cite{Hilaire12}
\end{itemize}

These models have proven their capacity to reproduce relatively accurately and systematically experimental data available despite the relatively different energy-, spin- and parity-dependent descriptions of the NLD  \cite{Capote09}.
For a systematic comparison with Oslo data, all models are treated on the same footing. More specifically, they are renormalized to reproduce as well as possible  the cumulative number of low-lying levels and the $s$-wave resonance spacings (see Sec.~\ref{sec:oslo}). To do so, the same procedure as in Ref.~\cite{Goriely08b} is followed, {\it i.e.} the level density $\rho_{th}$ is corrected by the expression
\begin{equation}
 \tilde\rho_{th}(U,J,\pi)=e^{\alpha\sqrt{(U-\delta)}} \times \rho_{th}(U-\delta,J,\pi) 
 \label{eq:renorm2}
 \end{equation}
 where $U$ is the excitation energy. The excitation-energy shift $\delta$ is essentially extracted from the analysis of the cumulative number of discrete levels, while $\alpha$ is determined from the experimental $s$-wave neutron spacing $D_{0}$. 
 With such a renormalization, the experimental low-lying
 states and the $D_{0}$ values can be reproduced reasonably well as discussed in detail in Refs.~\cite{Koning08,Goriely08b}.
However, in the specific case of the Cst-T model where the energy dependence of $\ln \rho_{th}$ is proportional to $U$ and not to $\sqrt{U}$, like in all the other models considered here, 
the following expression 
\begin{equation}
 \tilde\rho_{\mathrm{th,Cst-T}}(U,J,\pi)=e^{\alpha(U-\delta)} \times \rho_{th}(U-\delta,J,\pi) 
  \label{eq:renorm1}
 \end{equation}
is preferred to ensure the constant-temperature behavior is kept at low energies. 
Through Eq.~(\ref{eq:renorm1}), the usual pairing shift and temperature in the Cst-T formula are adjusted through the $\delta$  and $\alpha$ parameters, respectively, to ensure a proper description of the low-lying levels and $D_0$ value.
 
\subsection{Extraction of NLD from the Oslo method}
\label{sec:oslo}
Here we give a brief description of the main idea and analysis steps in the Oslo method. 
For more details, we refer the reader to Refs.~\cite{Guttormsen87,Guttormsen96,Schiller00,Larsen11}.
The probability of $\gamma$ decay from an initial excitation-energy bin $E_i$ to a final excitation-energy bin $E_f$ by a  $\gamma$ ray of energy $E_\gamma = E_i-E_f$  is proportional
to the level density at the final excitation-energy bin, $\rho(E_f)$, and the $\gamma$-ray transmission coefficient $\mathcal{T}(E_\gamma)$. 
Hence, the experimental  first-generation $\gamma$-ray matrix can be factorized
into~\cite{Schiller00}
\begin{equation}
P(E_\gamma,E_i) \propto \mathcal{T}(E_\gamma) \rho(E_i-E_\gamma),
\label{eq:P}
\end{equation}
where the bin-wise normalization $\sum_{E_\gamma} P(E_\gamma,E_i) = 1$ has been applied.
This factorization is justified when the nucleus reaches a compound state before de-excitation, which ensures that the manner of the
subsequent $\gamma$ decay is mainly statistical and independent of how the state was formed.
The $\gamma$-transmission coefficient ${\cal T}$ is a function of $E_\gamma$ only, in accordance with
the Brink hypothesis~\cite{Brink55}, which in its generalized form states that any collective decay mode has the same properties whether it is built on the ground state or on excited states.

Unique NLD and GSF can, however, not be extracted from the first-generation $\gamma$-ray matrix. In particular, as shown in Ref.~\cite{Schiller00}, the NLD obtained by the transformation 
\begin{equation}
\tilde\rho(E_i-E_\gamma)=\rho(E_i-E_\gamma)Ae^{\alpha^\prime (E_i-E_\gamma)}
\label{eq:rhoslo}
\end{equation}
fits equally well to the experimental data, provided that the $\gamma$-transmission coefficient is also transformed by the term $e^{\alpha^\prime E_\gamma}$ with the same slope adjustment $\alpha'$:
\begin{equation}
\tilde{\cal{T}} (E_\gamma) = {\cal{T}}(E_\gamma)B e^{\alpha^\prime E_\gamma}.
\label{eq:tr_oslo}
\end{equation}
Here, $A$ and $B$ are scaling parameters giving the absolute normalization of $\rho$ and ${\cal{T}}$, respectively.
Because of the transformations given in Eqs.~(\ref{eq:rhoslo}-\ref{eq:tr_oslo}), the NLD data are normalized to the known, discrete levels and the $s$-wave resonance spacing $D_0$ at the neutron separation  
energies derived from the analysis of resolved resonances in low-energy neutron capture experiments~\cite{Capote09,Mughabghab06}. 
Thus the $A$ and $\alpha'$ parameters are determined, while the $B$ parameter is found using the average radiative width of the neutron resonances.

It should be noted that, even in the cases where $D_0$ values are available, the normalization procedure of the NLD introduces two model-dependent uncertainties \cite{Larsen11,Goriely19}. 
These are: (\textit{i}) the assumption of parity equilibrium, {\it i.e.}, as many negative-parity and positive-parity levels at the neutron 
 separation energy, and (\textit{ii}) that the total NLD at the neutron 
 separation energy can be calculated from the weighting of the $D_0$ values with the spin distribution given by Refs.~\cite{Bethe36,Ericson58}.
Regarding the first assumption, this is likely fulfilled for most nuclei with $A\ga 40$, as shown both theoretically  (e.g. the HFB+comb calculations of Ref.~\cite{Goriely08b}, with typical parity distributions shown in Figs.~23-25 of Ref.~\cite{Larsen11}) and experimentally~\cite{Kalmykov07}.
The second assumption can potentially introduce a significant systematic uncertainty, and will be discussed in detail in the following. 

For $s$-wave neutron-resonance experiments where $I_t$ is the spin of the target nucleus ground state and $\pi_t$ its parity, the neutron resonance spacing $D_0$ can be written in terms of the partial level density for the involved spin(s) and parity as
\begin{equation}
\frac{1}{D_0}=  \sum_{J_f} \rho(B_n,J_f=|I_t \pm \frac{1}{2}|,\pi_t).
\label{eq:d0}
\end{equation}
Specificallly, for a target nucleus with $I_t = 0^+$ capturing an $s$-wave neutron with eigenspin $s=1/2$ and  orbital angular momentum $\ell = 0$, the populated levels in the compound nucleus will have final spin $J_f = 1/2$ and positive parity. 
If $I_t>0$, the levels populated in the capture process have spins $J_f = I_t \pm 1/2$ with positive parity if  $\pi_t = +$, or negative parity if $\pi_t = -$.
Relation~(\ref{eq:d0}) is justified by the fact that all levels with $J_f = | I_t \pm 1/2 |$ are accessible in an $s$-wave neutron resonance experiment.

The total NLD at $B_n$, $\rho(B_n)$, is found by combining Eq.~(\ref{eq:d0}) with the spin-dependent NLD. 
In the Cst-T, BSFG and GSM approaches (Sec.~\ref{sec:mod}), one assumes equiparity and that the spin dependence can be expressed as $\rho(U,J)=\rho(U) \cdot g(U,J)$, where $\rho(U)$ is the total NLD at excitation energy $U$, and $g(U,J)$ is the spin distribution given by~\cite{Bethe36,Ericson58} 
\begin{equation}
g(U,J) \simeq \frac{2J+1}{2\sigma^2(U)}\exp\left[-(J+1/2)^2/2\sigma^2(U)\right],
\label{eq:spindist}
\end{equation}
where $\sigma(U)$ is the spin cutoff parameter that is  excitation-energy dependent.
Then, the total NLD at $B_n$ can be deduced by
\begin{equation}
\rho(B_n)=\frac{2}{D_0}\frac{1}{\sum_{J_f} g(B_n,J_f)},
\label{eq:rhobn}
\end{equation}
where the factor of 2 comes from the assumption of equiparity at $B_n$.
When the spin-dependence cannot be so easily separated, e.g. in the HFB+comb or THFB+comb models, the total NLD can still be deduced from the sum of the NLD at all spins, provided Eq.~(\ref{eq:d0}) is fulfilled when applied to  the spins $J_f=I_t \pm 1/2$ only. 
Finally, note that equiparity is assumed in all NLD models, except in the HFB+comb and THFB+comb models.

\subsection{Renormalisation procedure}
\label{sec:ren}

As mentioned above, in the standard Oslo method the slope of the NLD is not experimentally constrained but still subject to the unknown $e^{\alpha^\prime (E_i-E_\gamma)}$ factor (see Eq.~\ref{eq:rhoslo}). 
For this reason, the measured NLD data are traditionally renormalized to the low-lying levels and the total NLD at the neutron 
 separation energy deduced from the $D_0$ value for a given NLD model (Eq.~\ref{eq:d0}), as described in Sec.~\ref{sec:oslo}. 
As far as the extrapolation between the highest energy data points and the neutron
 separation energy is concerned, most of the previous analyses have been performed assuming an exponential character of the energy dependence, or in other words using a Cst-T formula (in some studies, a BSFG was also used). 
To be consistent, the same NLD model should, however, be used for both the determination of $\rho(B_n)$ and the extrapolation at the highest energy data points. 
So, in this work, in contrast to most of the previous analyses, the NLD extrapolation from $B_n$ down to the experimental NLD point at the highest excitation energy $E_{\exp}^{\rm max}$ is not performed by the Cst-T or BSFG formulae, but rather consistently by the actual NLD model in question (adjusted on the low-lying levels and $D_0$ value).
As usually done in the previous Oslo-data analyses, several data points at high excitation energies are to be considered in the normalization; in practice we consider the 20 data points below $E_{\exp}^{\rm max}$. 
Finally, the normalization corresponds to the lowest possible root-mean-square (rms) deviation with the model in question, defined as 
\begin{equation}
f_{rms}= \mathrm{exp} \left[\frac{1}{N_e} \sum_{i=1}^{N_e} \ln^2 r_i \right]^{1/2}  
\label{eq:frms}
\end{equation} 
where $N_e$ is the number of data points included for a given nucleus and $r_i$ is, for each data point $i$, the ratio of the theoretical to experimental
level density which takes into account the experimental uncertainties $\delta \rho_{\rm exp}$ affecting $\rho_{\rm exp}$, as follows:
\begin{eqnarray}
r=&&\frac{\rho_{th}}{\rho_{{\rm exp}} - \delta\rho_{{\rm exp}}} \quad {\rm if} \quad \rho_{th} < \rho_{{\rm exp}} - \delta\rho_{{\rm exp}} \nonumber \\
=&&\frac{\rho_{th}}{\rho_{{\rm exp}} + \delta\rho_{{\rm exp}}}\quad {\rm if} \quad \rho_{th} > \rho_{{\rm exp}} + \delta\rho_{{\rm exp}} \nonumber \\
=&&1 \quad \hskip 2.1cm{\rm otherwise.}  
\label{eq_rat}
\end{eqnarray}


\begin{figure*}
\centering
\includegraphics[scale=0.5]{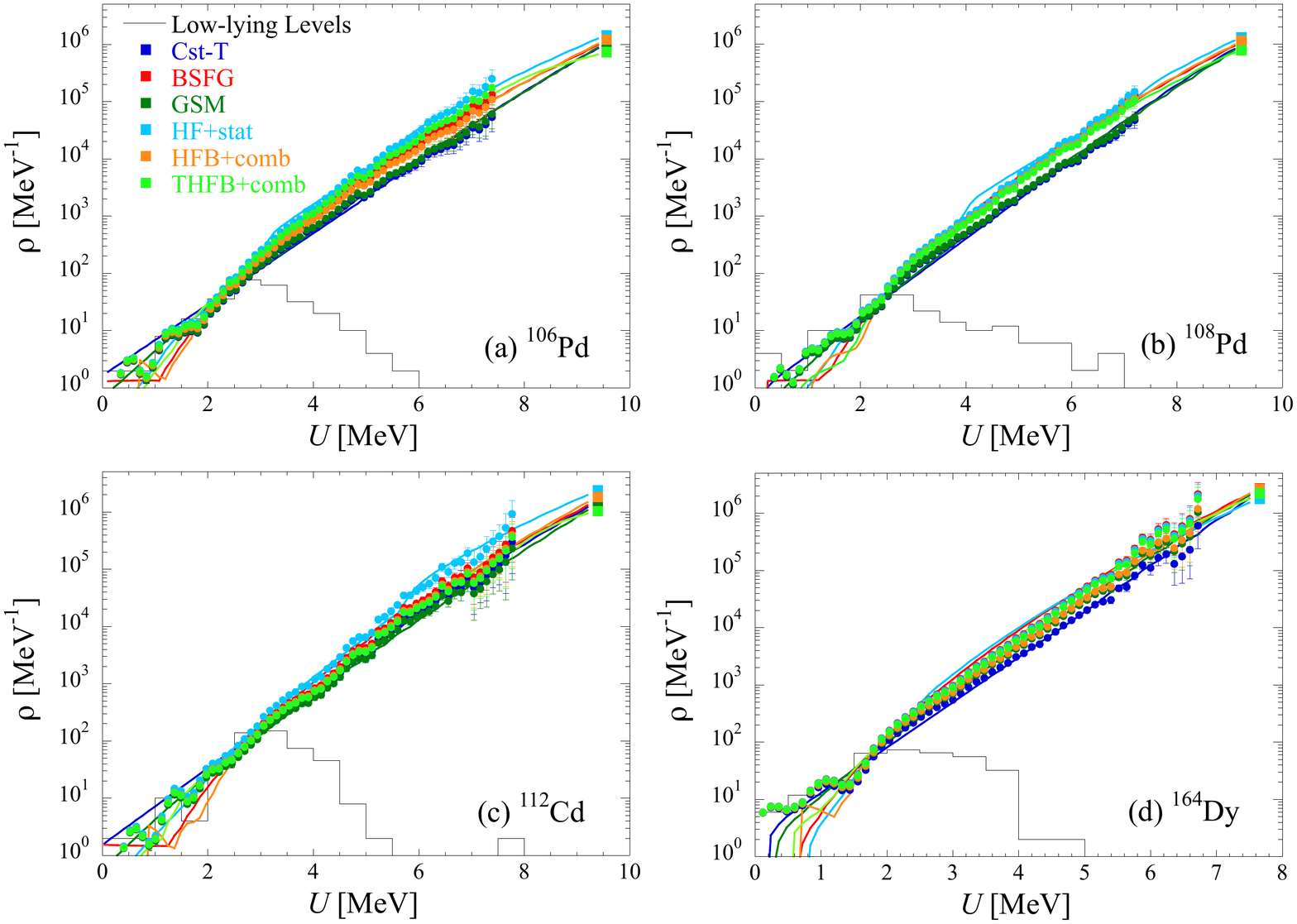}
\caption{(Color online) Comparison of 6 NLD predictions with Oslo data after renormalization for the four test cases (a) $^{106}$Pd, (b) $^{108}$Pd, (c) $^{112}$Cd and (d) $^{164}$Dy. 
The solid black line represents the NLD extracted from known discrete levels using an excitation-energy bin $\Delta U=0.5$~MeV.  
The filled circles correspond to the measured NLDs renormalized by Eq.~(\ref{eq:rhoslo}) on the corresponding theoretical NLD to minimize $f_{rms}$. 
The lines correspond to the six NLD calculations normalized to the experimental $D_0$ value. 
The full squares give the predicted total NLD at $U=B_n$ for each of the six NLD models.}
\label{fig_testcase}
\end{figure*}

In summary, for each NLD model, the following procedure is adopted:
\begin{enumerate}
\item the NLD formula is adjusted to match the low-lying levels through the $\delta$ parameter in Eq.~(\ref{eq:renorm2}) (or Eq.~\ref{eq:renorm1} in the case of the Cst-T formula). The energy at which the level scheme is assumed to be complete ($E_{lls}$) is estimated from a comparison with Oslo data;
\item the NLD formula is normalized to the $D_0$ experimental value through the $\alpha$ parameter in Eq.~(\ref{eq:renorm2}) (or Eq.~\ref{eq:renorm1} in the case of the Cst-T formula);
\item the total NLD model at the neutron separation energy $\rho_{th}(B_n)$ is deduced;
\item the total NLD model as a function of excitation energy $\rho_{th}(U)$ is used to extrapolate the NLD at the highest excitation energies measured with the Oslo technique;
\item the Oslo level densities and errors at the highest excitation energies are normalized through the parameter $\alpha^\prime$ in Eq.~(\ref{eq:rhoslo}). The $f_{rms}$ deviation between the NLD formula and the renormalized Oslo measurements is calculated, 
and the parameter $\alpha^\prime$ deduced from the minimum $f_{rms}$ value;
\item Steps 1-5 are re-iterated a few times to ensure that $\alpha$, $\delta$ and $\alpha^\prime$ parameters optimize the reproduction of the low-lying levels by the Oslo data, and at the same time keeping the constraint on $D_0$; 
\item the rms deviation between the NLD prediction and the renormalized Oslo measurements is estimated for all the excitation-energy points included in the Oslo measurement.
\end{enumerate}
Each NLD model described in Sec.~\ref{sec:mod} is treated on the same footing, except the Cst-T formula for which Eq.~(\ref{eq:renorm1}) replaces Eq.~(\ref{eq:renorm2}). 
The same procedure has been applied to the 42 nuclei for which Oslo measurements have been performed \cite{Oslo} and for which $s$-wave resonance spacings are experimentally known \cite{Capote09,Mughabghab06}.

\section{Test cases: $^{106,108}$Pd, $^{112}$Cd and $^{164}$Dy}
\label{sec:test}

Four even-even nuclei  are considered in the present section and their NLDs studied in detail. 
They correspond to $^{106,108}$Pd, $^{112}$Cd and $^{164}$Dy.  The result of the normalization procedure given in Sec.~\ref{sec:ren} is illustrated in Fig.~\ref{fig_testcase}.

\begin{figure}
\includegraphics[scale=0.42]{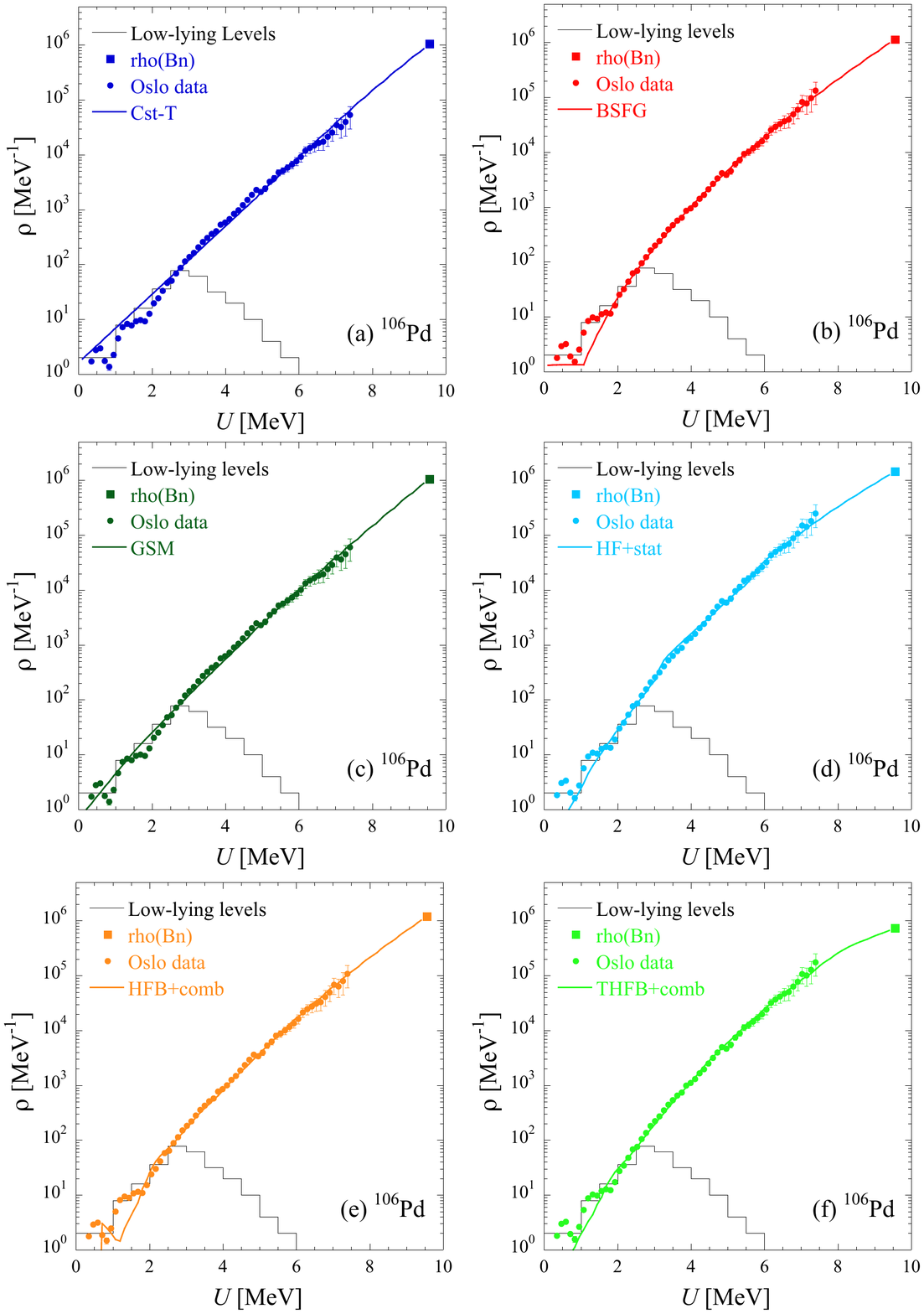}
\caption{(Color online) Theoretical and renormalized Oslo NLD for $^{106}$Pd for each of the six NLD models considered. The black solid  line represents the NLD extracted from known discrete levels using an excitation-energy bin of $\Delta U=0.5$~MeV.  
The filled circles correspond to the measured NLD renormalized by Eq.~(\ref{eq:rhoslo}) to the theoretical NLD at $E_{\exp}^{\rm max} \simeq 7.4$~MeV. The additional lines correspond to six NLD calculations (Sec.~\ref{sec:mod}). The full squares at $U=B_n$ give the total NLD extracted for a given NLD model after renormalization to the same experimental $D_0$ value. }
\label{fig_106pd_ldx}
\end{figure}

The $^{106}$Pd NLD has been measured by the Oslo method and presented in Ref.~\cite{Eriksen14} where the extrapolation between the total NLD estimated from the $D_0$ value and the upper Oslo data points was performed using the BSFG model. 
The $s$-wave average spacing is estimated to be $D_0=10.3 \pm 0.5$~eV \cite{Capote09} obtained from the neutron capture of the $I_t^\pi=5/2^+$ $^{105}$Pd target with a neutron separation energy $B_n=9.561$~MeV. 
The different NLD predictions constrained on low-lying levels and $D_0$ are shown in the upper left panel of Fig.~\ref{fig_testcase}. 
The BSFG is found to give the best description of the NLD shape deduced from Oslo data, as confirmed by the $f_{rms}$ values given in Table~\ref{tab:frms}. 
The combinatorial models give 
almost a similar accuracy. 
The model with the worst agreement in this case is the Cst-T formula. 
Interestingly, at the highest excitation energies, the renormalized Oslo NLDs may differ by a factor 4.7 between the largest HF+stat  and the lowest Cst-T predictions. 
To highlight the agreement obtained with each NLD model, the six models are shown separately in Fig.~\ref{fig_106pd_ldx}.

Similar results can be found for $^{108}$Pd, $^{112}$Cd and $^{164}$Dy. 
At medium-to-high excitation energies ($\approx 3-7$ MeV or so), the HF+stat models usually give the largest NLD predictions, while the Cst-T or GSM give the lowest ones. 
However, this conclusion is not necessarily valid for the NLD at the neutron separation energy $B_n$, as seen in Fig.~\ref{fig_testcase}.

\begin{table}
\begin{center}
\caption {$f_{rms}$ deviations for the four test nuclei calculated on the $\sim 40$ energy points above the energy $E_{lls}$ where the set of low-lying levels is not longer complete, {\it i.e.} when the Oslo NLD becomes larger than the NLD deduced from low-lying levels. }
  \begin{tabular}{ lcccc}
  \hline
NLD model	&	$^{106}$Pd & $^{108}$Pd & $^{112}$Cd & $^{164}$Dy 	\\
  \hline
  \hline
  Cst-T		&	1.10	&	1.17	&	1.04	&	1.13	\\
BSFG		&	1.02	&	1.10	&	1.09	&	1.17	\\
GSM			&	1.07	&	1.13	&	1.05	&	1.09	\\
HF+stat		&	1.08	&	1.22	&	1.09	&	1.30	\\
HFB+comb	&	1.04	&	1.11	&	1.05	&	1.08	\\
THFB+comb	&	1.04	&	1.12	&	1.05	&	1.09	\\
    \hline
 \end{tabular}
\label{tab:frms}
\end{center}
\end{table}

\section{Systematic comparison}
\label{sec:sys}

\begin{figure*}
\centering
\includegraphics[scale=0.68]{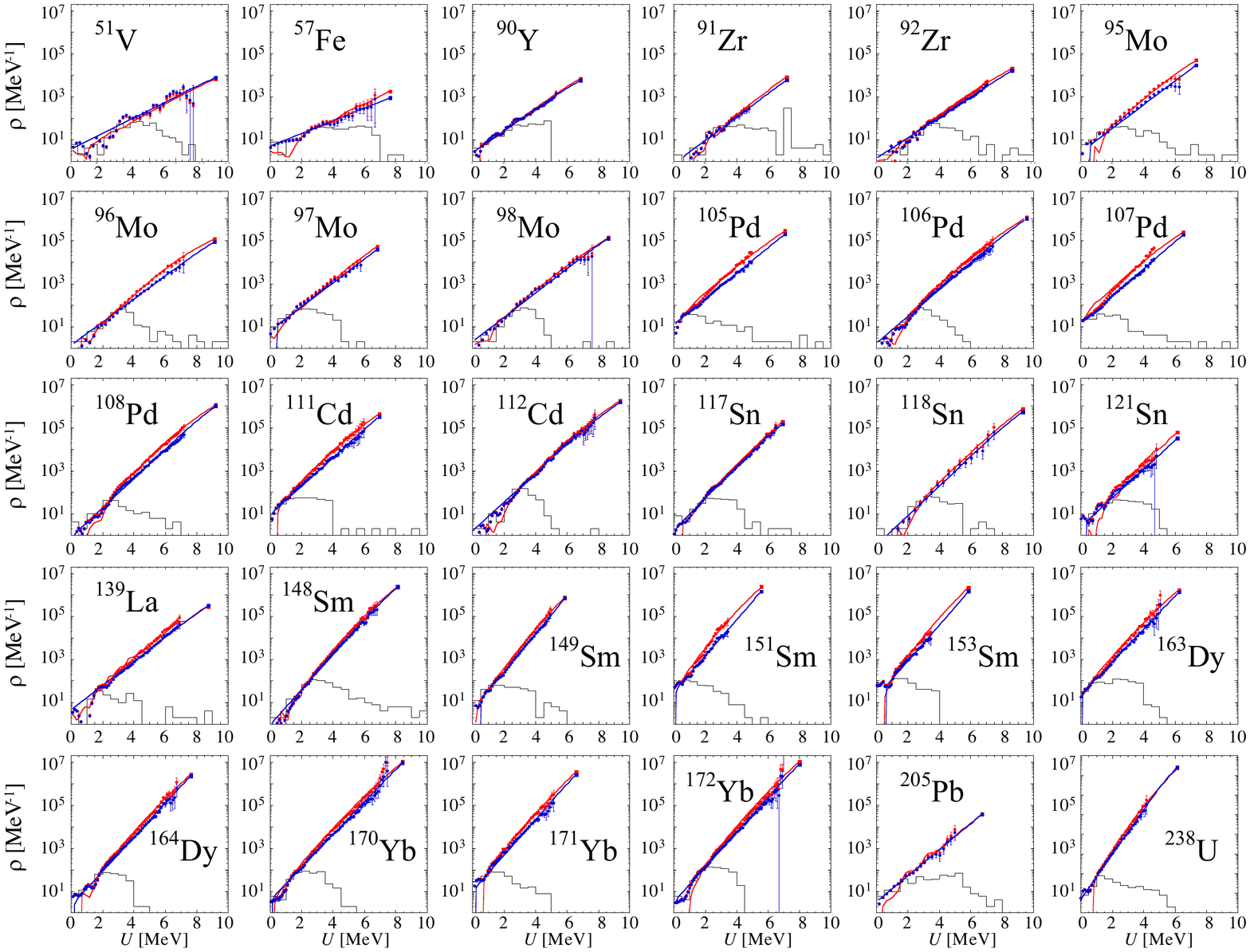}
\caption{(Color online) Comparison for 30 nuclei between the renormalized experimental Oslo data (filled circles) and NLD curves (solid lines) corresponding to the Cst-T (blue) or HFB+comb (red) models. The black solid lines are the NLDs deduced from the low-lying level schemes.}
\label{fig_sys_ld1-5}
\end{figure*}

The test cases discussed in Sec.~\ref{sec:test} are now extended more systematically to the 42 nuclei for which Oslo measurements \cite{Oslo} and a $D_0$ value  \cite{Capote09} are available. 
Fig.~\ref{fig_sys_ld1-5} compares in particular the NLD predictions and experimental Oslo data renormalized with the procedure detailed in Sec.~\ref{sec:ren} for 30 nuclei out of the 42 when considering the Cst-T or HFB+comb models. 

\begin{figure}
\centering
\includegraphics[scale=0.3]{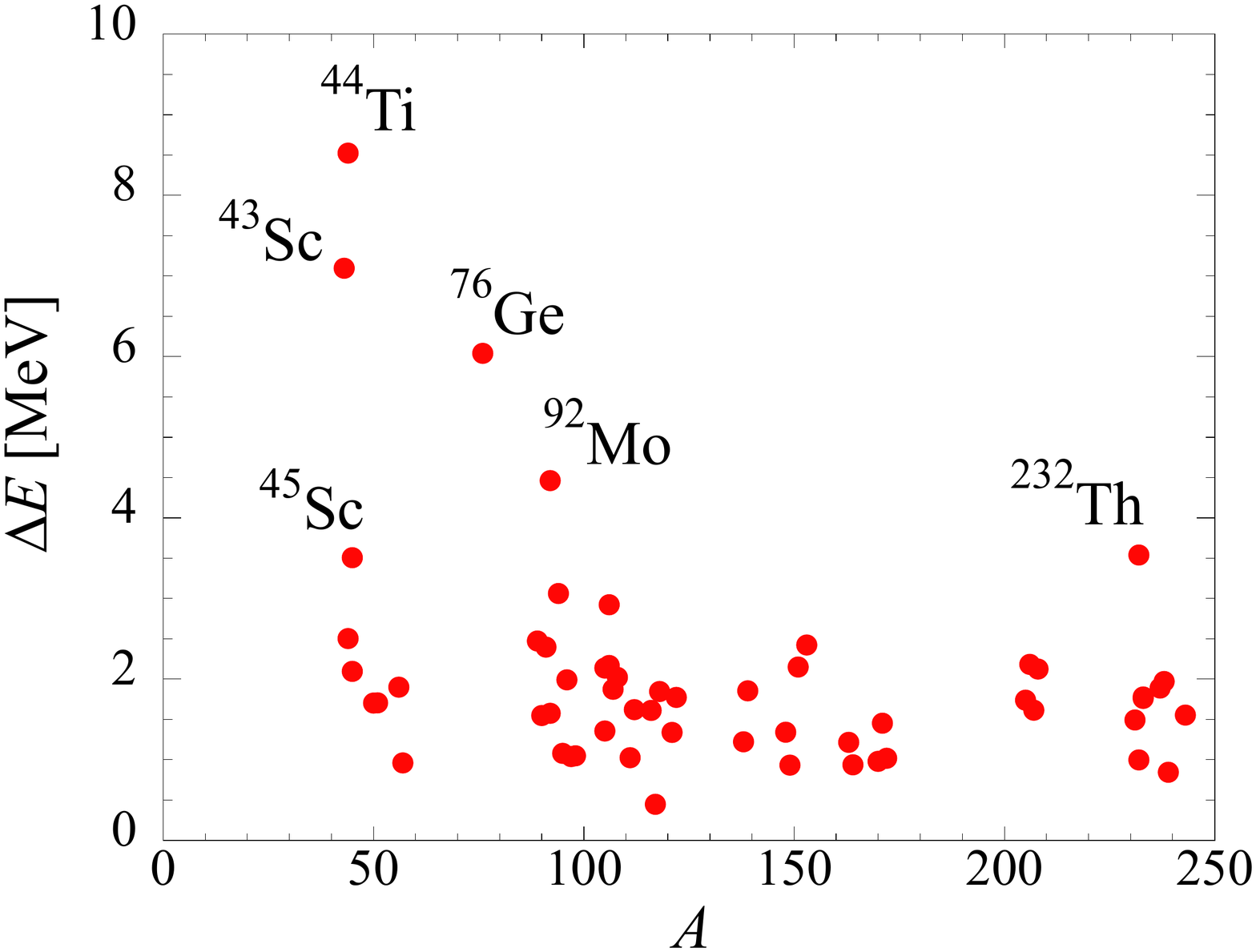}
\caption{(Color online) Energy difference $\Delta E=B_n-E_{\exp}^{\rm max}$ between the neutron separation energy and the data point at the highest excitation energy  as a function of the atomic mass $A$ for all nuclei for which Oslo measurements are available. The nuclei with the largest energy differences ($\Delta E > 3.5$~MeV) are marked specifically. }
\label{fig_emax}
\end{figure}

Globally, a rather good agreement is found with both NLD models (see e.g. $^{90}$Y). 
However, in some cases, significant deviations between both models and the corresponding renormalized Oslo data can be observed, either due to different $D_0$ values predicted (e.g. in the case of $^{95}$Mo), or different shapes of the NLD excitation-energy dependence (e.g. for $^{108}$Pd), or both (see, for example, $^{151}$Sm). 
Naturally, the larger the energy difference $\Delta E=B_n-E_{\exp}^{\rm max}$ between the data point at the highest excitation energy  $E_{\exp}^{\rm max}$ and $B_n$, the larger the impact of the NLD model on the normalization procedure. 
We show in Fig.~\ref{fig_emax} this energy difference $\Delta E$ for all nuclei for which Oslo measurements are available. 
The energy difference $\Delta E$ is found to lie mainly between 1 and 2~MeV, but can reach values as high as 8~MeV (for $^{44}$Ti). 
In particular, six nuclei are found to have a $\Delta E > 3.5$~MeV and in addition no measured $D_0$ value. 
In this case, the extraction of the experimental NLD from the Oslo method remains highly challenging and significantly more affected by model-dependent uncertainties.

\begin{table}
\begin{center}
\caption {Mean $\varepsilon$ and rms $\sigma$ deviations for all the $N_n=42$ nuclei corresponding, for a given NLD model, to the differences between the NLD predictions and the newly renormalized Oslo data. The first two columns include all Oslo data points, the last two columns only include energy points above the energy $E_{lls}$.
} 
  \begin{tabular}{ lcccc}
  \hline
    NLD model  & $\varepsilon$(all) & $\sigma$(all) & $\varepsilon (E>E_{lls})$ & $\sigma (E>E_{lls})$ \\
    \hline
    \hline
Cst-T		&	1.02	&	1.45	&	0.97	&	1.21	\\
BSFG		&	0.92	&	1.68	&	1.01	&	1.25	\\
GSM			&	0.97	&	1.69	&	1.00	&	1.34	\\
HF+stat		&	0.94	&	1.53	&	1.02	&	1.27	\\
HFB+comb	&	0.94	&	1.47	&	0.99	&	1.25	\\
THFB+comb	&	0.95	&	1.64	&	1.02	&	1.30	\\    
\hline
 \end{tabular}
\label{tab:rms}
\end{center}
\end{table}

Out of these systematic calculations, it seems possible to evaluate quantitatively which models reproduce at best the shape of the experimental data. 
 To do this, we define a mean and an rms deviation over all the $N_n=42$ nuclei for which Oslo measurements and $s$-wave resonance spacings are available, as
\begin{eqnarray}
& & \varepsilon= \mathrm{exp} \left[ \frac{1}{N_n}\sum_{n=1}^{N_n} \left(  \frac{1}{N_e} \sum_{i=1}^{N_e} \ln \frac{\rho^i_{th}}{\rho^i_{exp}} \right) \right] \\
& &\sigma = \mathrm{exp} \left[\frac{1}{N_n}\sum_{n=1}^{N_n} \left( \frac{1}{N_e} \sum_{i=1}^{N_e} \ln^2 \frac{\rho^i_{th}}{\rho^i_{exp}}\right)  \right]^{1/2}  ~.
\label{eq:sig}
\end{eqnarray} 
The deviations are given in Table~\ref{tab:rms}. 
As the conclusions may depend on the adopted set of data points, we consider here two options. 
The first one includes \textit{all} Oslo data points, the second only includes data points at energies above an excitation energy $E_{lls}$, where the set of low-lying levels is not longer complete, {\it i.e.} when the Oslo NLD becomes larger than the NLD deduced from low-lying levels.

When considering all data points available, the Cst-T and HFB+comb models come out as the models describing the experimental data with the highest accuracy, i.e. the lowest rms deviation of the order of 1.45 corresponding to a global ratio of about 45\% around the mean value  (the link between the $f_{rms}$ value and the data description can also be visualized by comparing Table~\ref{tab:frms} and Figs.~\ref{fig_testcase}-\ref{fig_106pd_ldx}). A slightly larger deviation, {\it i.e.} lower accuracy, is obtained with the HF+stat model.
Larger discrepancies up to a factor of about 1.6-1.7 are obtained with the THFB+comb, BSFG and GSM models, which, after renormalization, are therefore about 20\% less accurate than the Cst-T or HFB+comb models to describe the energy dependence of the NLD extracted from the Oslo data. All models present a mean value close to one, {\it i.e.} no global overestimation or underestimation.\\
If we only consider excitation energies above $E_{lls}$, the Cst-T model remains the most precise one, though all models are seen to give rise to a rather similar accuracy with an rms deviation of $\sigma\sim 1.2-1.3$.
Comparing rms deviations for both sets, {\it i.e.} the full data set and only energies above $E_{lls}$, it can be deduced that, at the lowest energies, NLDs are less accurately described by the BSFG, GSM and THFB+comb models in comparison with the other models. All rms deviations are clearly reduced when omitting the energies below $E_{lls}$, showing that it remains complex to describe the low-lying levels with an independent-particle model, regardless of its statistical or combinatorial nature.
In conclusion, this analysis shows that the Cst-T,  HFB+comb and HF+stat models present the best description of the Oslo data but also that it remains difficult to favour one of these models from such a global analysis.

\section{Constraints by the Shape Method}
\label{sec:shape}

The recently developed shape method~\cite{Wiedeking21} has been used \cite{Mucher22} to provide the absolute measured NLD of even-even nuclei $^{76}$Ge and $^{88}$Kr. In this analysis, the slope parameter $\alpha'$ of the Oslo GSF is extracted from a $\chi^2$ comparison with the shape method GSF.
Hence, only the scaling parameter $A$ in Eq.~(\ref{eq:rhoslo}) remains to be determined by normalizing the NLD to the known discrete levels. This leads to a much reduced uncertainty associated with the renormalization procedure inherent to the Oslo method \cite{Mucher22}. We here apply this technique to the case of $^{112}$Cd.

\begin{figure}[tb]
 \begin{center}
 \includegraphics[clip,width=1\columnwidth]{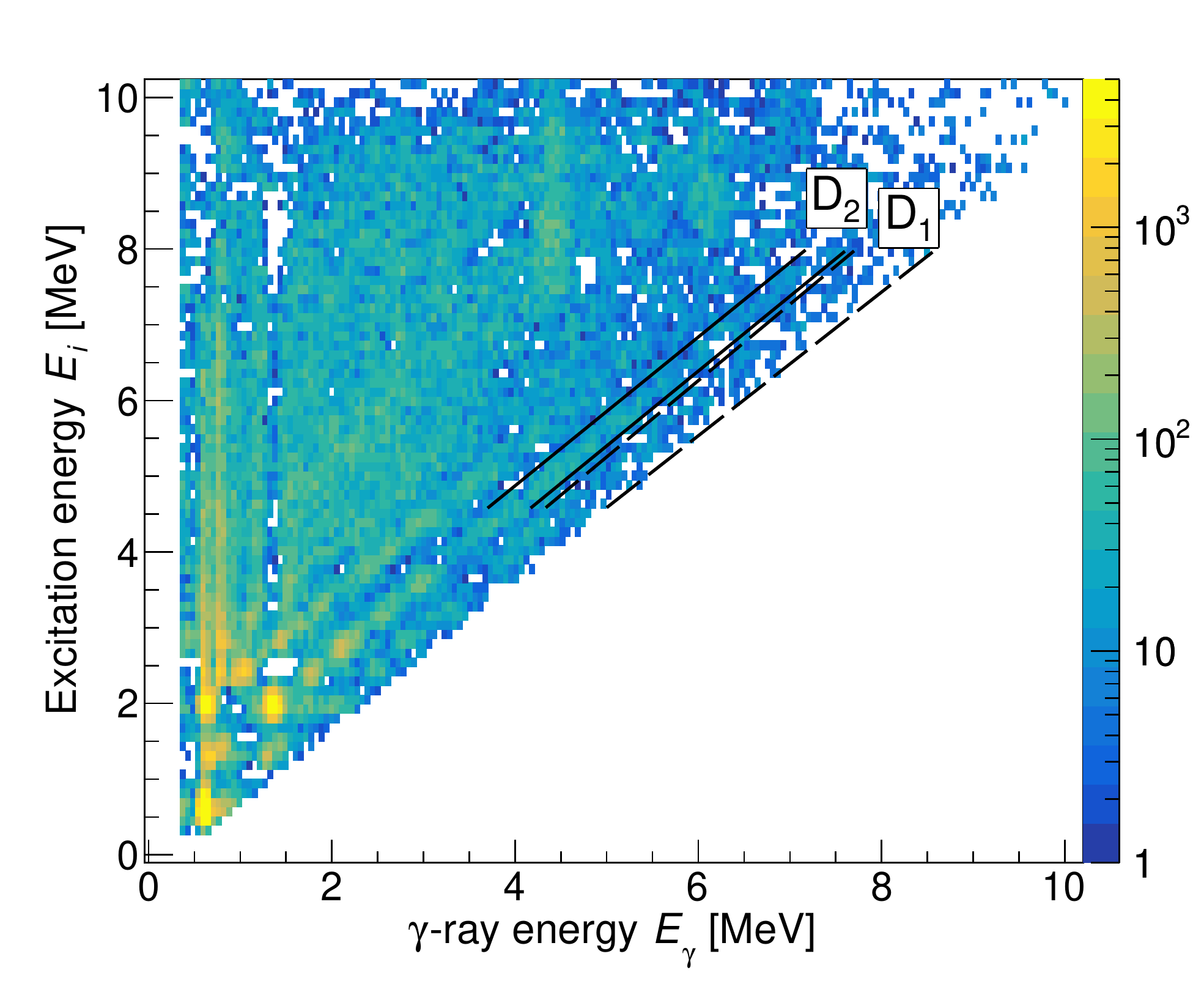}
 \caption {(Color online) The first-generation matrix of $^{112}$Cd. The black dashed lines indicate the limits for the decay to the ground state ($D_1$) and the  black solid lines the decay to the first $2^+$ ($D_2$). }
 \label{fig:fgmatrix}
 \end{center}
 \end{figure}

The basic idea of the shape method is that the \textit{functional form} of the GSF can be determined directly from the intensity (number of counts) of the $\gamma$ decay from highly excited states to specific discrete states, such as the ground state and the first excited state. 
Indeed, the GSF 
\begin{equation}
f^{i}_{XL}(E_\gamma) = \frac{{\cal T}_{XL}(E_\gamma)}{2\pi E_\gamma^{2L+1}} = \frac{\left< \Gamma_{\gamma i}(E_\gamma) \right>}{E_\gamma^{2L+1}D_i}
\end{equation} 
is a measure of the reduced partial radiative width $\Gamma_{\gamma i}$ \cite{Bartholomew73} for transitions from an initial excitation-energy bin $i$ with energy $E_\gamma$, electromagnetic character $X$, multipolarity $L$, and average level spacing $D_i$.
For our analysis, we assume that the average strength function is independent of the initial energy, spin and parity, and that the dipole $L=1$ transitions dominate \cite{Brink55,Larsen13}.
We also only consider decay to the ground state $0^+$ and the first $2^+$, with the corresponding number of counts $N_{0^+}$ and $N_{2^+}$, respectively. 
This is done by choosing appropriate limits for the diagonals in the first-generation matrix that correspond to the decay to the ground state $0^+$ (diagonal $D_1$) and the first $2^+$ (diagonal $D_2$), taking into consideration the detector resolution, and then simply integrating the number of counts $N_{0^+}$ and $N_{2^+}$ for each excitation-energy bin $E_i$. 
The first-generation matrix $P(E_\gamma,E_i)$ of $^{112}$Cd with the applied limits for the diagonals is shown in Fig.~\ref{fig:fgmatrix}.
 \begin{figure*}[tb]
 \begin{center}
 \includegraphics[clip,width=2\columnwidth]{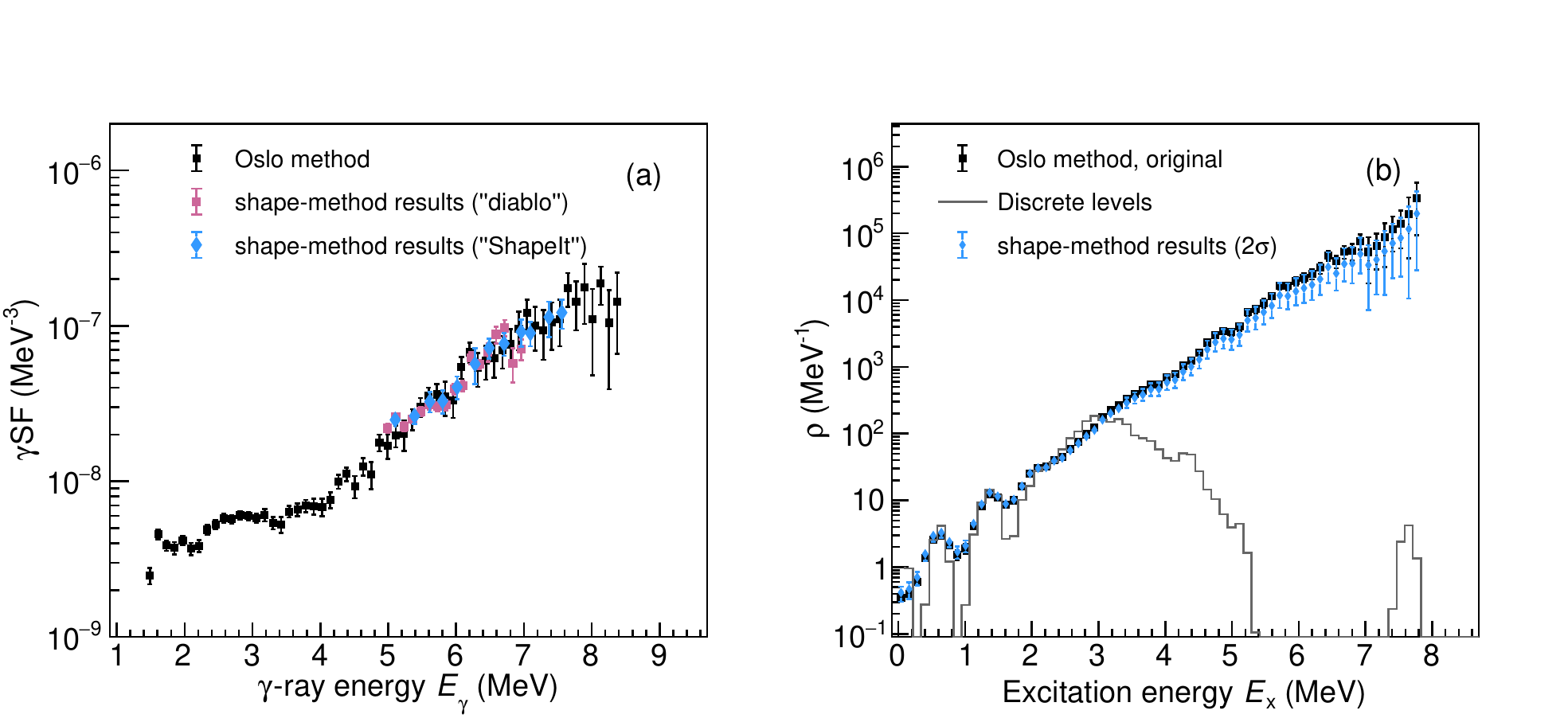}
 \caption {(Color online) (a) Extracted GSFs for $^{112}$Cd using the Oslo method (black squares) and the shape method with two different codes (pink squares and light blue diamonds);  (b) the original NLD from the Oslo method (black squares), and the renormalized one using the slope from the shape method (light blue diamonds). The error bars on the shape-method renormalized NLD include statistical errors and represent a $2\sigma$ confidence level. }
 \label{fig:rsfnldcomp}
 \end{center}
 \end{figure*}
 \begin{figure}[tb]
 \begin{center}
 \includegraphics[clip,width=1\columnwidth]{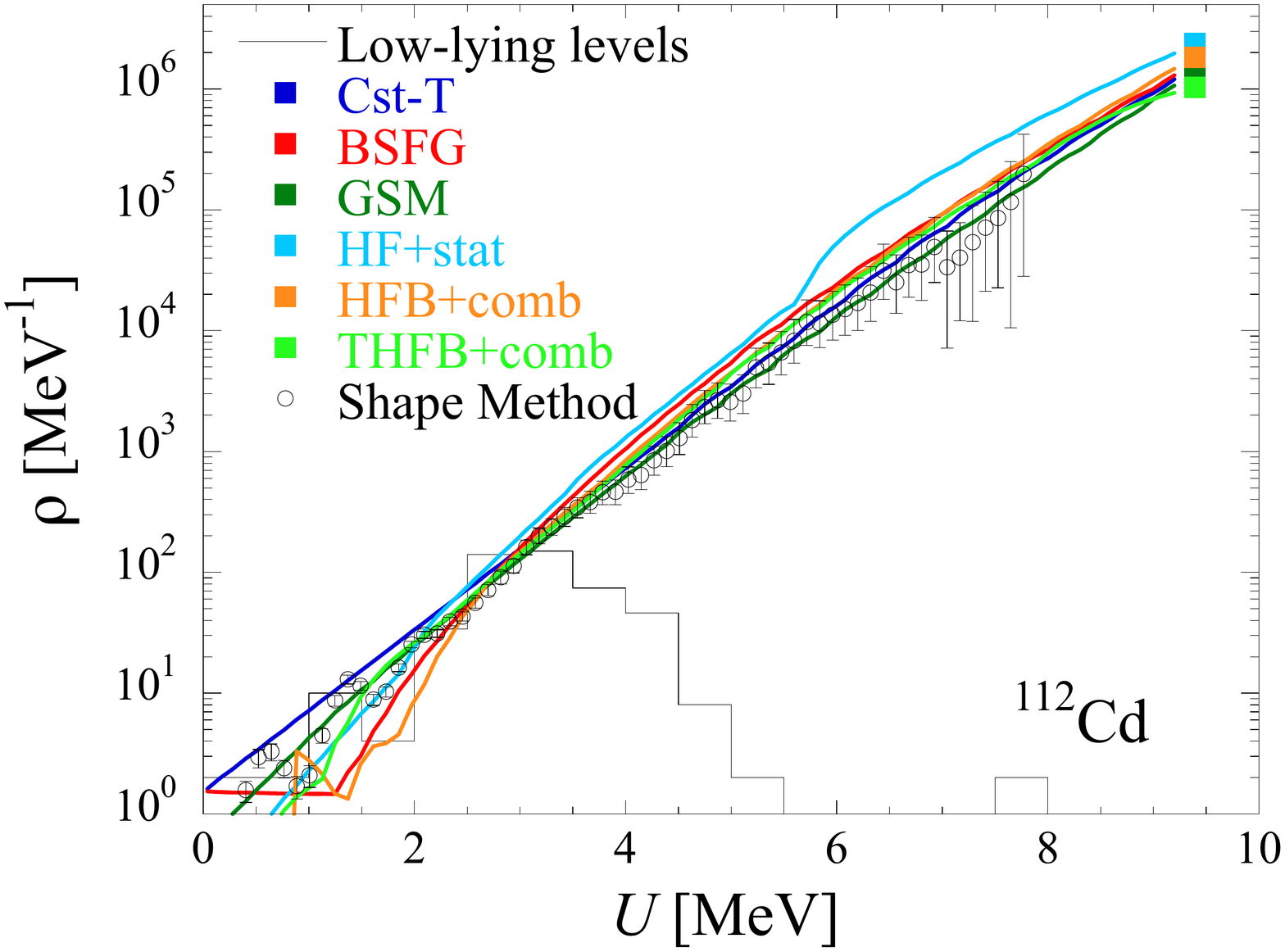}
 \caption {(Color online) Comparison, for $^{112}$Cd, between the NLD constrained by the shape method (open circles) and the NLD predicted by the six models considered here (solid lines). The full squares at $U=B_n$ give the total NLD extracted for a given NLD model after renormalization to the same experimental $D_0$ value. The black solid  line represents the NLD extracted from known discrete levels using an excitation-energy bin of $\Delta U=0.5$~MeV. }
 \label{fig:shapecomp}
 \end{center}
 \end{figure}

We can relate the number of counts to the GSF for a given excitation-energy bin $E_i$ by
\begin{equation}
N_{0^+}(E_i) = \eta f(E_\gamma) E_\gamma^3, 
\end{equation}
and 
\begin{equation}
N_{2^+}(E_i) = \eta f(E_\gamma) E_\gamma^3 \frac{g(J_i=1)}{\left[ g(J_i=1) + g(J_i=2) + g(J_i=3)\right]},
\end{equation}
where $g(J_i)$ is the spin distribution of the initial states with spin $J_i$ and $\eta$ is a scaling factor that depends on the population cross section for the given initial excitation-energy bin.
The factor 
\begin{equation}
R = g(J_i=1)/\left[ g(J_i=1) + g(J_i=2) + g(J_i=3)\right]
\label{eq:spinratio}
\end{equation} accounts for the fact that we have only one possible $J_i$ for the ground state, namely $J_i = 1$ (for dipole transitions), while, for the $2^+$, three possibilities exist, namely $J_i = 1,2,3$. 
As discussed previously, the spin distribution is model dependent if the spin cutoff parameter is unknown experimentally ({\it i.e.} in almost all cases). 
However, different models for the spin cutoff parameter give nearly the same ratio $R$ for the case of $^{112}$Cd for the range of initial excitation energies used here ($E_i = 5.5-8.0$ MeV), which means that the shape method is not much affected by this uncertainty (at least in this particular case). 
For example, using the spin cutoff parameter from Ref.~\cite{Egidy05,Egidy06}, $R$ varies from 0.23 at $E_i = 5.5$ MeV to 0.22 at $E_i = B_n$.  

As the factor $\eta$ is unknown, we employ a sewing technique, normalizing each pair $N_{0^+}, N_{2^+}$ for a given $E_i$ internally by a logarithmic interpolation, as described in Ref.~\cite{Wiedeking21}, using the code \textsf{diablo} available at the Oslo-method software Github~\cite{OsloGithub}. 
Here, the explicit variation of the spin cutoff parameter as a function of the excitation energy is included.
In addition, we have used the code \textsf{ShapeIt}~\cite{Mucher22}, where a slightly different sewing technique is implemented, and where uncertainties, e.g. due to the choice of bin size, are taken into account.
Note that in performing such an internal normalization, the method relies on the Brink hypothesis to be valid for adjacent $E_i$ bins. 
Finally, we obtain a GSF that has a fixed slope together with  uncertainties due to statistical and possible systematic errors in the sewing technique such as a possible dependency on the bin size. 
To compare this shape-method GSF to the Oslo-method GSF, we perform a $\chi^2$ minimization of the two data sets.
This procedure provides an absolute scaling of the shape-method GSF, so that we can perform a new $\chi^2$ test to see whether the slope of the Oslo data corresponds to the slope of the shape-method results. 

The resulting GSFs are shown in Fig.~\ref{fig:rsfnldcomp}a, where we have extracted shape-method results with both codes \textsf{diablo} and \textsf{ShapeIt}. 
We observe a good agreement with the original Oslo data, but the $\chi^2$ test reveals that the slope (see Eq.~\ref{eq:tr_oslo}) should be corrected with a factor $\exp(\Delta\alpha' E_\gamma)$ where $\Delta\alpha' = -0.1$ to have the optimal agreement between the shape-method and Oslo-method GSFs. 
This implies the same slope correction on the NLD according to Eqs.~(\ref{eq:rhoslo}-\ref{eq:tr_oslo}), as shown in Fig.~\ref{fig:rsfnldcomp}b. 
Here, the error bars in the slope-corrected NLD data represent a $2\sigma$ confidence limit from the $\chi^2$ test. 
By such an analysis, combined with the normalization to the known discrete levels, we can put strong constraints on the \textit{absolute} NLD without any resort to the neutron resonance spacings and associated renormalization procedure. 

The comparison for $^{112}$Cd between the final shape-method NLD and our six NLD models is shown in Fig.~\ref{fig:shapecomp}.
As can be seen, the models that overall reproduce best the data are the Cst-T and the GSM models in this case. The HF+stat and the BSFG models have the largest discrepancies, in particular the HF+stat model for excitation energies $U=6-8$ MeV.

Of course, there are limitations to the shape method since its precision depends on several factors: 
\begin{itemize}
    \item[1.] {to what degree the Brink hypothesis is fulfilled, as this hypothesis must be invoked for the internal normalization (sewing technique) of the shape-method data.}
    \item[2.] {how high is the initial level density (partial level density)$-$a low level density can lead to large Porter-Thomas fluctuations, making the internal normalization uncertain~\cite{Markova22b}.}
    \item[3.] {for cases where the spins of the final levels are not the same, a spin distribution must be applied to estimate the ratio $R$ in Eq.~(\ref{eq:spinratio}).  This spin distribution is in general not known experimentally and could induce a systematic error.}  
\end{itemize}
These potentially hampering factors should be considered for each individual case for which the shape method is applied. 
For $^{112}$Cd, we restrict ourselves to high initial excitation energies, $E_i = 5.5-8.0$ MeV, which ensures a high partial NLD ($\rho(E_i=5.5\mathrm{MeV},J_i=1) \sim 600$ MeV$^{-1}$).
With such a high NLD, the first two above-mentioned points should not cause major problems. 
Point 3 is already discussed and found not to be a significant factor in the case of $^{112}$Cd.
In conclusion, in agreement with the earlier work of Ref.~\cite{Mucher22}, we have here demonstrated that the shape method, in combination with the Oslo method, can provide an \textit{absolute} NLD for $^{112}$Cd that is essentially model-independent, as long as the above-mentioned factors are carefully considered. Reducing statistical and systematic uncertainties will be key in the future to constrain the energy dependence of the NLD. Applying such a method in the future to a large sample of nuclei can help us evaluating the quality of the energy dependence proposed by the different models available and consequently predicting NLD for experimentally unknown nuclei. 

\section{Conclusions}
\label{sec:con}
For the last two decades, experimental NLD have been obtained on the basis of the Oslo method \cite{Guttormsen87,Guttormsen96,Schiller00,Oslo}, consisting today in data for about 60 different nuclei. While each of these measurements has been individually compared to one or a few NLD models, a global study including all the measurements has been missing. Such an analysis can provide insight on the energy dependence of the NLD below the neutron separation energy and on the validity of the NLD models for a large range of nuclei at excitation energies below the neutron
 separation energy. 
Since the NLD extracted from the Oslo method still needs to be renormalized, a coherent well-defined model-dependent procedure has been applied and six different NLD models, all being treated on the same footing, systematically compared with Oslo data for the 42 nuclei for which $s$-wave spacings at the neutron 
separation energy are available.

 Our  quantitative analysis shows that the constant-temperature model presents the best global description of the Oslo data, closely followed by the mean-field plus combinatorial models and Hartree-Fock plus statistical model. However, their accuracies remain rather similar, so that it remains difficult to clearly favour one of these models. The other models still perform but with a smaller quality in the data description. When considering energies above the threshold where the experimental level scheme is complete, all the six models are shown to give rather similar accuracies.
These models are characterized by a different energy dependence that cannot be differentiated at this stage due to the unavoidable model-dependent nature of the renormalization procedure applied to the Oslo data. It remains hard to exclude some of models considered here as long as the renormalization procedure is casting doubt on the exact slope of the experimental NLD. In this respect, the newly proposed shape method is shown to be promising since it can provide an absolute estimate of the energy dependence of the measured NLD and consequently reduces the uncertainties associated with the renormalization procedure inherent to the Oslo method. We have shown in the specific case of $^{112}$Cd that the shape method could exclude the HF+stat model and favour the Cst-T model. Such an analysis remains to be performed for the bulk of data for which Oslo measurements are available before evaluating the global quality of a given NLD model. Reducing the uncertainties associated with the shape method, in particular for the highest energy points, could also further increase the constraints on NLD models, not only regarding the energy dependence, but also potentially the spin dependence when coupled to the information stemming from $s$-wave spacings at the neutron separation energy.

\begin{acknowledgments}
S.G. acknowledges financial support from FNRS (Belgium). This work was partially supported by the Fonds de la Recherche Scientifique - FNRS and the Fonds Wetenschappelijk Onderzoek - Vlaanderen (FWO) under the EOS Project No O022818F and O000422F. 
A.C.L. gratefully acknowledges funding from the Research Council of Norway, project grant No. 316116. 
\end{acknowledgments}

\bibliographystyle{apsrev4-2}
\bibliography{astro}

\end{document}